\documentclass{moriond}

\def\Journal#1#2#3#4{{#1} {\bf #2}, #3 (#4)}
\usepackage{lineno}

\def\be{\begin{equation}}
\def\ee{\end{equation}}
\def\bea{\begin{eqnarray}}
\def\eea{\end{eqnarray}}

\newcommand{\Photo}{}

\begin{document}
\vspace*{4cm}
\title{RESULTS ON HEAVY ION COLLISIONS AT LHCB}

\author{ Yanxi ZHANG on behalf of the LHCb collaboration}

\address{ LAL, Universit\'e Paris-Sud, CNRS/IN2P3, Orsay, France}
%\linenumbers
\maketitle\abstracts{
    Heavy flavor production is important in heavy ion collisions to study both cold and hot
    nuclear matter effects. 
    The LHCb experiment can make unique contribution to heavy ion physics, owing to the full particle identification
    of the detector in the forward region and the ability to collect fixed target data with proton or lead beams. 
    This report describes recent results with proton-lead collision data collected in 2013 and the prospect of
    heavy-ion studies at LHCb.
}
\section{Introduction}
At high energy densities, hadrons are transformed into quark gluon plasma (QGP), a new state of matter in which quarks and gluons are
deconfined and move like freely. The hot QGP medium is expected to be produced in high energy heavy nucleus-nucleus
collisions. The formation and properties of the QGP can be studied with heavy flavor production. Heavy flavors are
produced via hard interactions at the early stage of the nucleus-nucleus collision,
so they will interact with the QGP when they traverse the medium later on. In nucleus-nucleus
collisions, the heavy flavor production is also affected by cold nuclear matter effects (CNM), which are present 
regardless of the formation of QGP. To disentangle the CNM effects from the QGP effects, heavy flavor production in
proton-nucleus collisions could be studied. 
In this report, we present the heavy-ion physics programs at LHCb and summarize the results on heavy flavor production studies in proton-lead data
collected by the LHCb detector.

\section{LHCb detector and data taking}
The LHCb detector is a single-arm forward spectrometer covering the pseudorapidity
range $2<\eta<5$. 
The LHCb experiment is designed for precision measurements in the $b$ and $c$ quark sectors, but it is becoming a general
purpose detector in the forward rapidity range. 
The LHCb detector consisting of the tracking systems, the hadron particle identification, calorimeters and the muon system,
is fully instrumented in the fiducial region. 
This feature makes it possible to study the heavy flavors in a unique
kinematic region, namely low transverse momentum $p_\mathrm{T}$, large rapidity $y$, very large or small Feynman $x_F$,
complementary to other LHC experiments.  The details of the design and performances of the LHCb detector could be found in the references~\cite{LHCb1,LHCb2}. 

So far the LHCb experiment has collected data of proton-proton ($pp$), proton-lead ($p$Pb) and lead-lead (PbPb) collisions, and also proton- or
lead-gas fixed target collisions thanks to the System for Measuring the Overlap with Gas (SMOG)~\cite{SMOG} at LHCb. 
The fixed target data will be discussed in
detail later. These data are collected at different discrete nucleon-nucleon center-of-mass energies, $\sqrt{s_\mathrm{NN}}$, from 54 GeV to 13
TeV. The analyses presented in this report are based on the $p$Pb data taken at $\sqrt{s_\mathrm{NN}}=5$ TeV in 2013. Since the
LHCb detector covers only one direction of the full acceptance, there are two distinctive beam configurations at LHCb for
the $p$Pb collisions. In the forward (backward) configuration, the proton (lead) beam enters LHCb detector from the
interaction point. 
The proton beam and the lead beam have different energies per nucleon in the laboratory frame, so the nucleon-nucleon center-of-mass frame is boosted in
the proton direction with a rapidity shift $\Delta y=0.46$. The LHCb acceptance for forward (backward) collision is
$1.5<y^*<4$ ($-5<y^*<-2.5$),\footnote{The rapidity $y^*$ is defined in the nucleon-nucleon rest frame with the momentum direction of proton as positive $z$-axis} and the
corresponding integrated luminosity is about $1.1$ nb$^{-1}$ ($0.5$ nb$^{-1}$) for the forward (backward) collisions.

\section{Prompt $D^0$ production}
\begin{figure}
\centering
\begin{minipage}{0.8\linewidth}
\centerline{\includegraphics[width=1.0\linewidth]{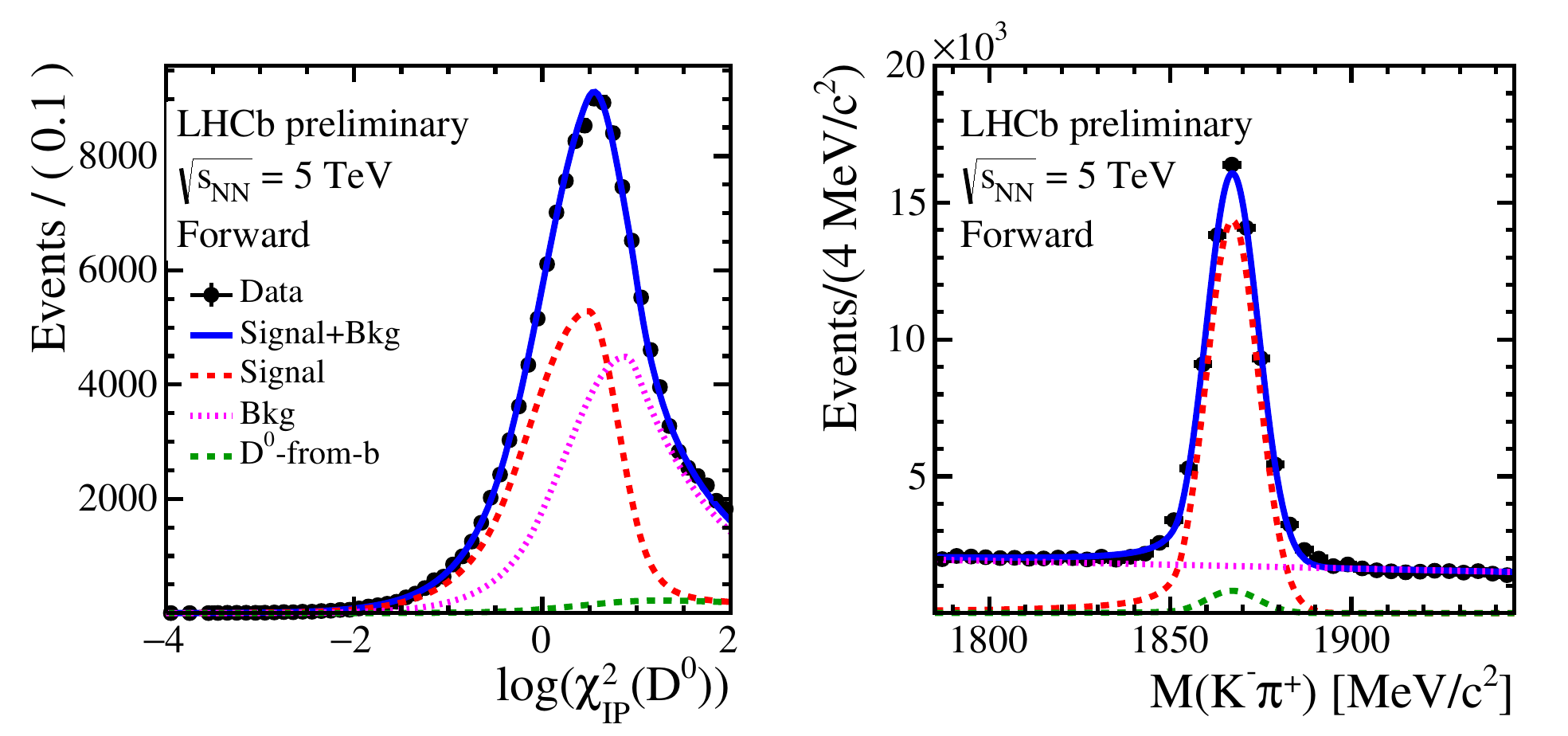}}
\end{minipage}
\caption[]{The distributions of (left) $\log(\chi^2_\mathrm{IP})$ and (right) invariant mass of the $D^0$ candidate in the
    forward sample. }
\label{fig:MassIPForward}
\end{figure}

At LHCb, the $D^0$ candidate is fully reconstructed in the $D^0\to K^-\pi^+$ decay mode, and selections are applied
exploring the large impact parameter and particle identification of the $K^-\pi^+$ tracks, and vertex displacement
of the $D^0$ candidate~\cite{LHCbD0}.  The signal and background events are discriminated by fitting the invariant mass
distribution, and the prompt $D^0$ and $D^0$ from $b$ hadron decays ($D^0$-from-$b$) are separated by fitting the impact
parameter significance ($\chi^2_\mathrm{IP}$) of the $D^0$ candidate.  The $\chi^2_\mathrm{IP}$ and invariant mass distribution for the forward
sample, integrating over all kinematic bins, are given in Fig.~\ref{fig:MassIPForward}. LHCb is the unique experiment
that could separate the $D^0$-from-$b$ component down to zero $p_\mathrm{T}$ region. For the cross-section measurements,
the reconstruction efficiency and particle identification efficiency are calibrated using data, and the distribution of
the number of tracks in simulation is corrected to match that in data.

\begin{figure}
\centering
\begin{minipage}{0.45\linewidth}
\centerline{\includegraphics[width=1.0\linewidth]{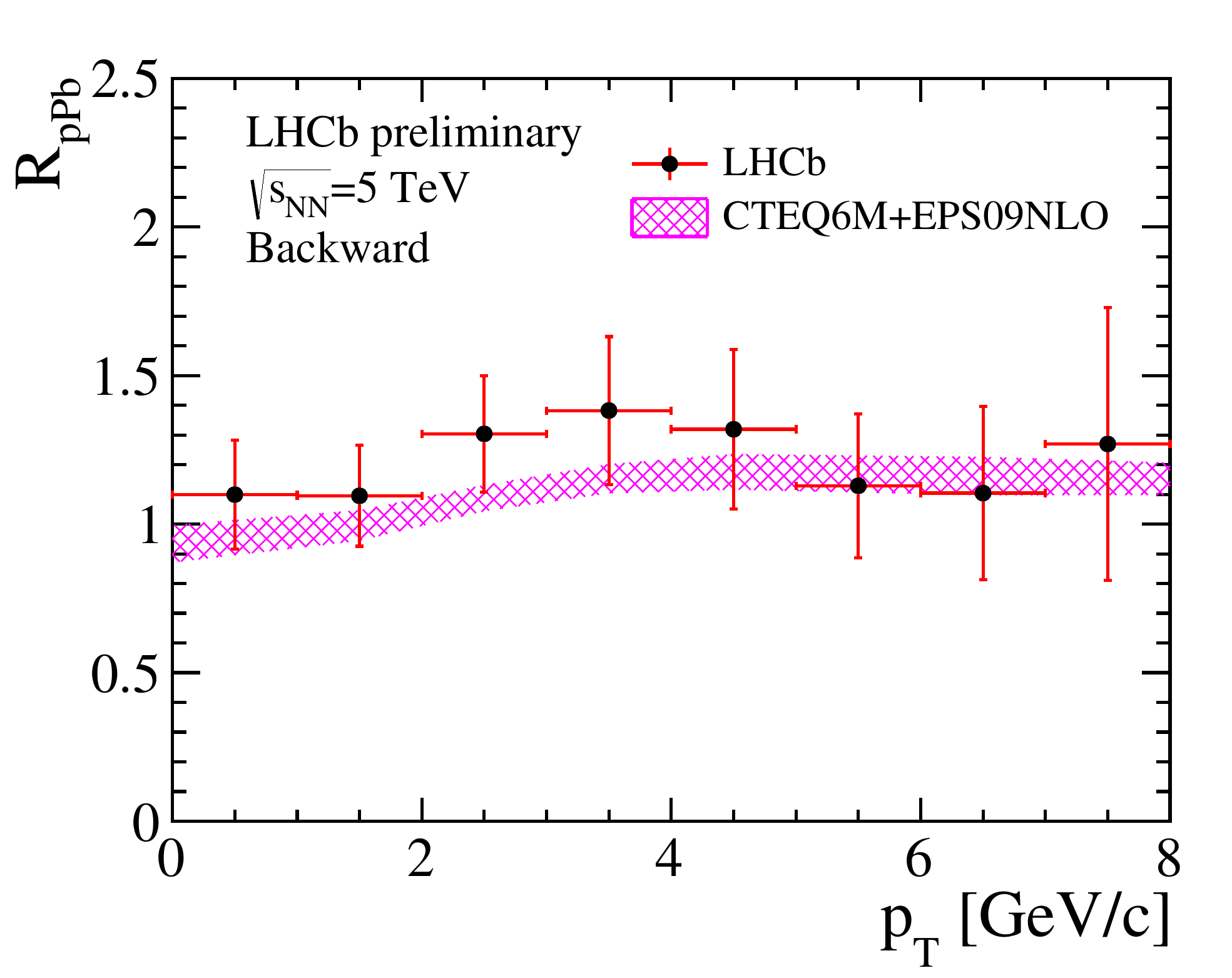}}
\end{minipage}
\hfill
\begin{minipage}{0.45\linewidth}
\centerline{\includegraphics[width=1.0\linewidth]{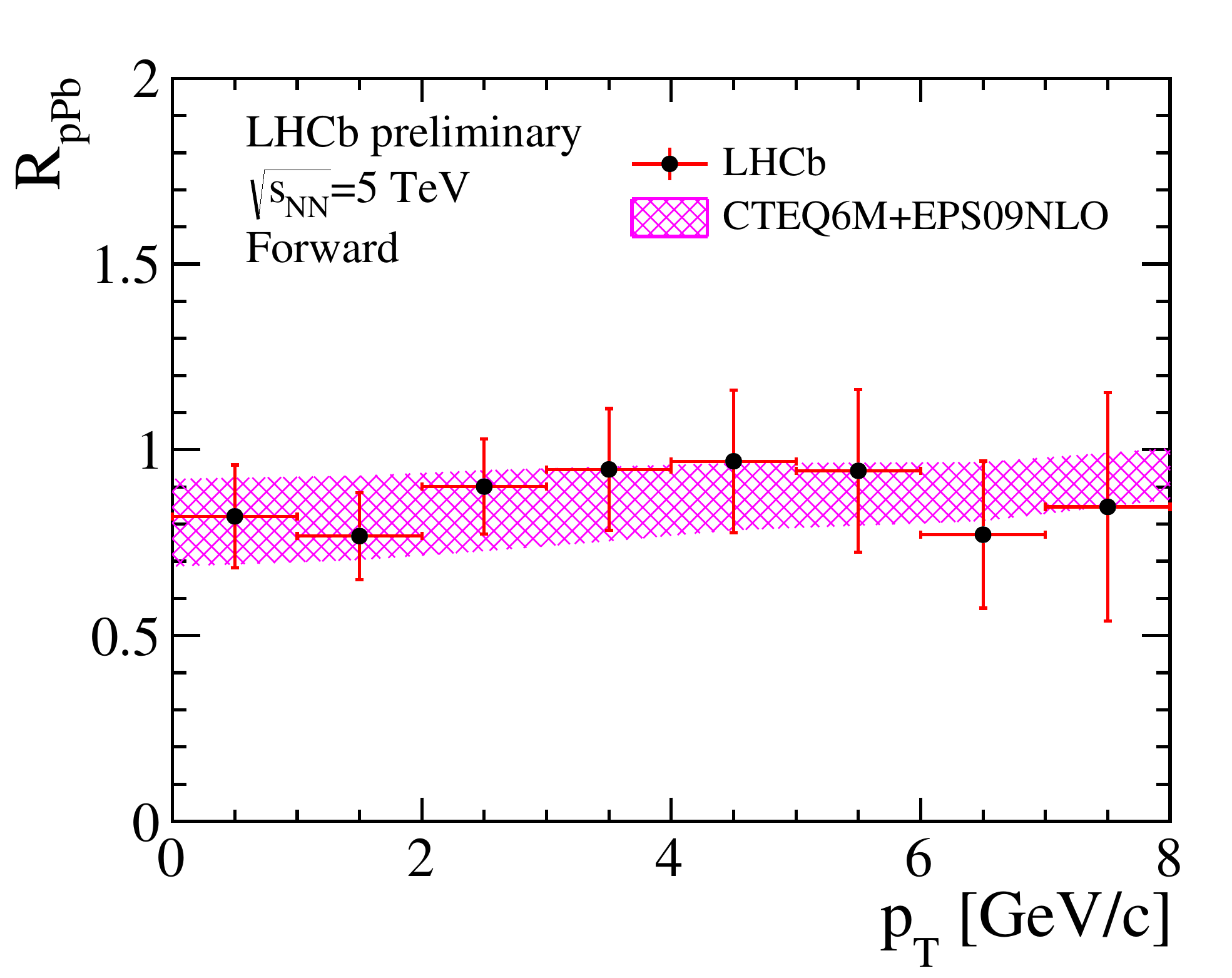}}
\end{minipage}
\caption[]{The nuclear modification factor $R_{p\mathrm{Pb}}$ as a function of $p_\mathrm{T}$ integrated over
    rapidity range $2.5<|y^*|< 4$ for (left) the backward sample and (right) the forward sample.}
\label{fig:RpPbPT}
\end{figure}

The double differential cross-section as a function of $p_\mathrm{T}$ and $y^*$ is measured in the range
$0< p_\mathrm{T}< 8$ GeV, and $1.5< y^*<4$ ($-5< y^*<-2.5$) for the forward (backward) sample. 
The nuclear modification factor, $R_{p\mathrm{Pb}}$, is determined to be the cross-section in $p$Pb collisions over that in $pp$
collisions at the same center-of-mass colliding energy, normalized by the number of binary colliding pairs $A$ ($A=208$ for $p$Pb collisions), as 
$R_{p\mathrm{Pb}} (y^*,p_\mathrm{T}) = \frac{1}{A}\times \frac{\sigma_{p\mathrm{Pb}}(y^*,p_\mathrm{T},\sqrt{s_\mathrm{NN}})}{\sigma_{pp}(y^*,p_\mathrm{T},\sqrt{s_\mathrm{NN}})}$.
The reference cross-section in $pp$ collisions at 5 TeV is determined by extrapolation from results at 7 and 13
TeV~\cite{D0At7TeV,D0At13TeV} with
the method described in reference~\cite{JpsiInPA}.
The reference cross-section measurements in $pp$ collisions at 5 TeV is underway.  
In Fig.~\ref{fig:RpPbPT}, the $R_{p\mathrm{Pb}}$ as a function of $p_\mathrm{T}$ integrated over the rapidity
range $2.5<|y^*|< 4$ are given for the backward (left) and forward (right) data. 
The large uncertainties are dominated by the extrapolated reference cross-section. Generally speaking, the nuclear
modification factor in the forward sample is smaller than that in the backward sample. Good agreements are
found between LHCb measurements and the MNR calculations~\cite{MNR} with CTEQ6M~\cite{CTEQ6M} and EPS09NLO~\cite{EPS09}
parton distribution functions.

\begin{figure}
\centering
\begin{minipage}{0.45\linewidth}
\centerline{\includegraphics[width=1.0\linewidth]{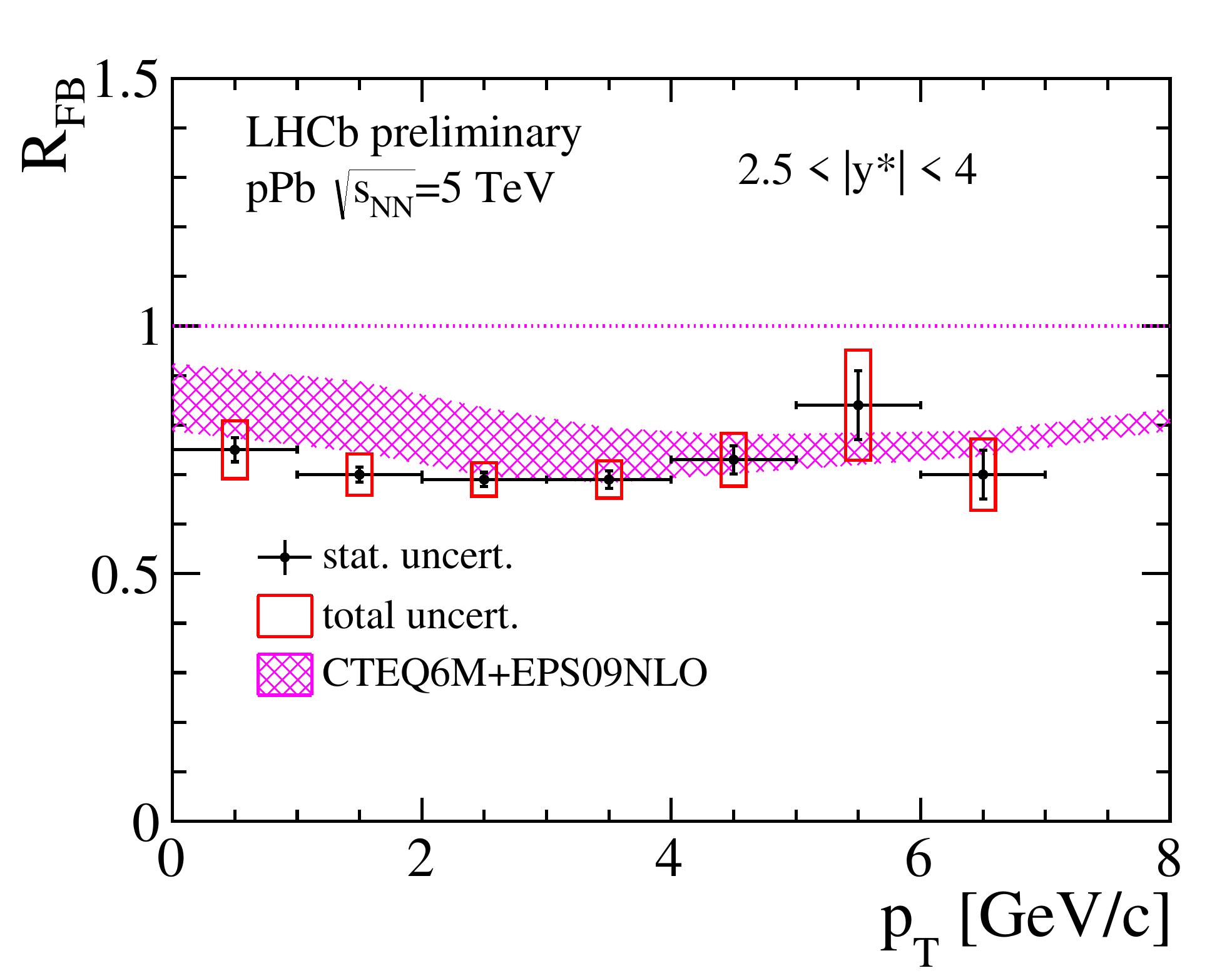}}
\end{minipage}
\hfill
\begin{minipage}{0.45\linewidth}
\centerline{\includegraphics[width=1.0\linewidth]{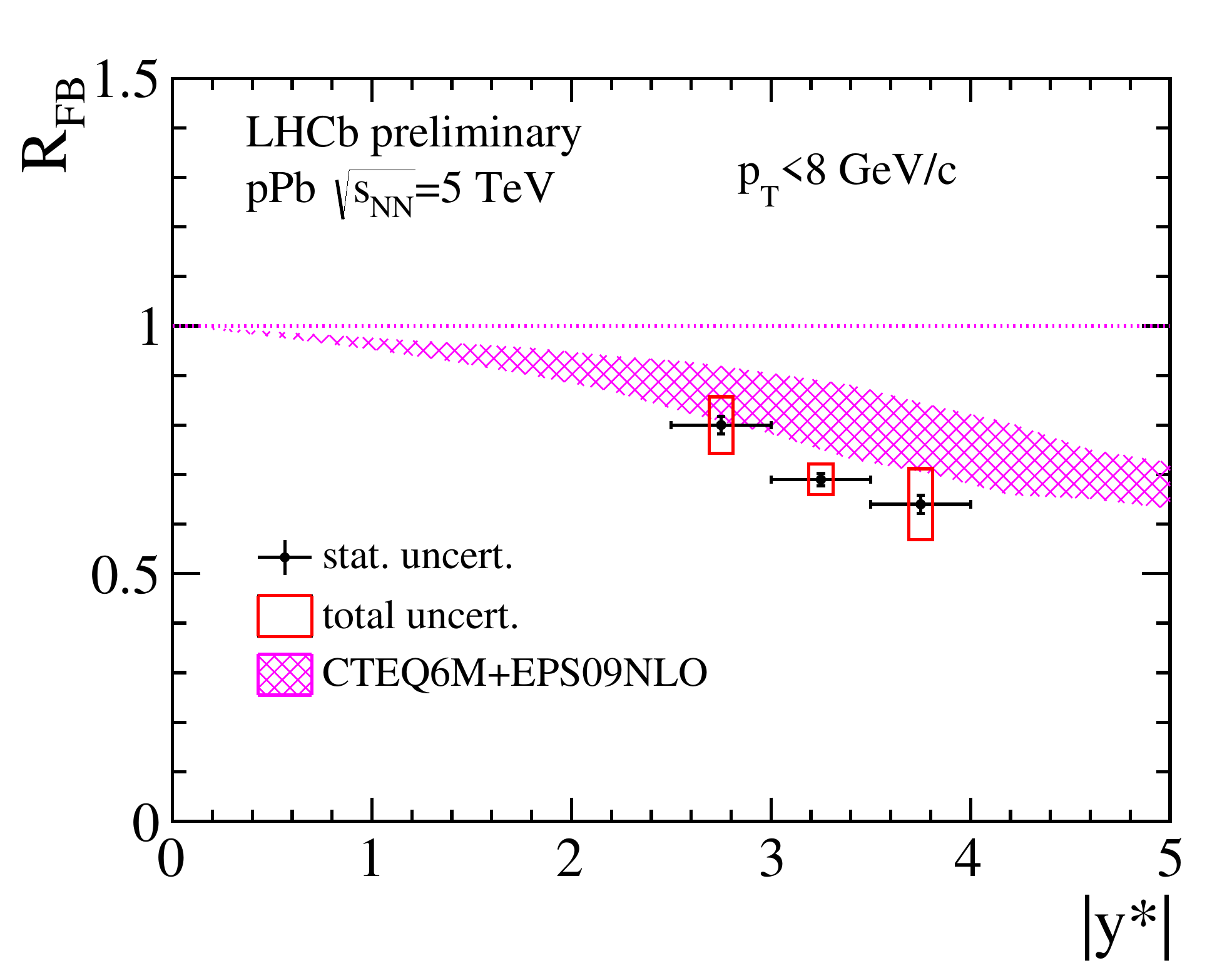}}
\end{minipage}
\caption[]{The forward-backward production ratio $R_\mathrm{FB}$ as a function of (left) $p_\mathrm{T}$ and (right)
    $y^*$. The integrated kinematic range is indicated also in the respective plots.}
\label{fig:RFB}
\end{figure}

The forward-backward ratio, $R_\mathrm{FB}$, is defined as
$R_\mathrm{FB}(|y^*|,p_\mathrm{T})=\frac{\sigma_{p\mathrm{Pb}}(+|y^*|,p_\mathrm{T})}{\sigma_{p\mathrm{Pb}}(-|y^*|,p_\mathrm{T})}$,
and systematic uncertainties largely cancel in the ratio. The $R_\mathrm{FB}$ as a function of $p_\mathrm{T}$ and $y^*$ is
given in Fig.~\ref{fig:RFB}, and the results suggest significant production asymmetry
between the forward and backward acceptance, indicating strong nuclear matter effects. The measurements are in
reasonable agreement with MNR calculations.

\section{Quarkonium productions}
\begin{figure}
\centering
\begin{minipage}{0.32\linewidth}
\centerline{\includegraphics[width=1.0\linewidth]{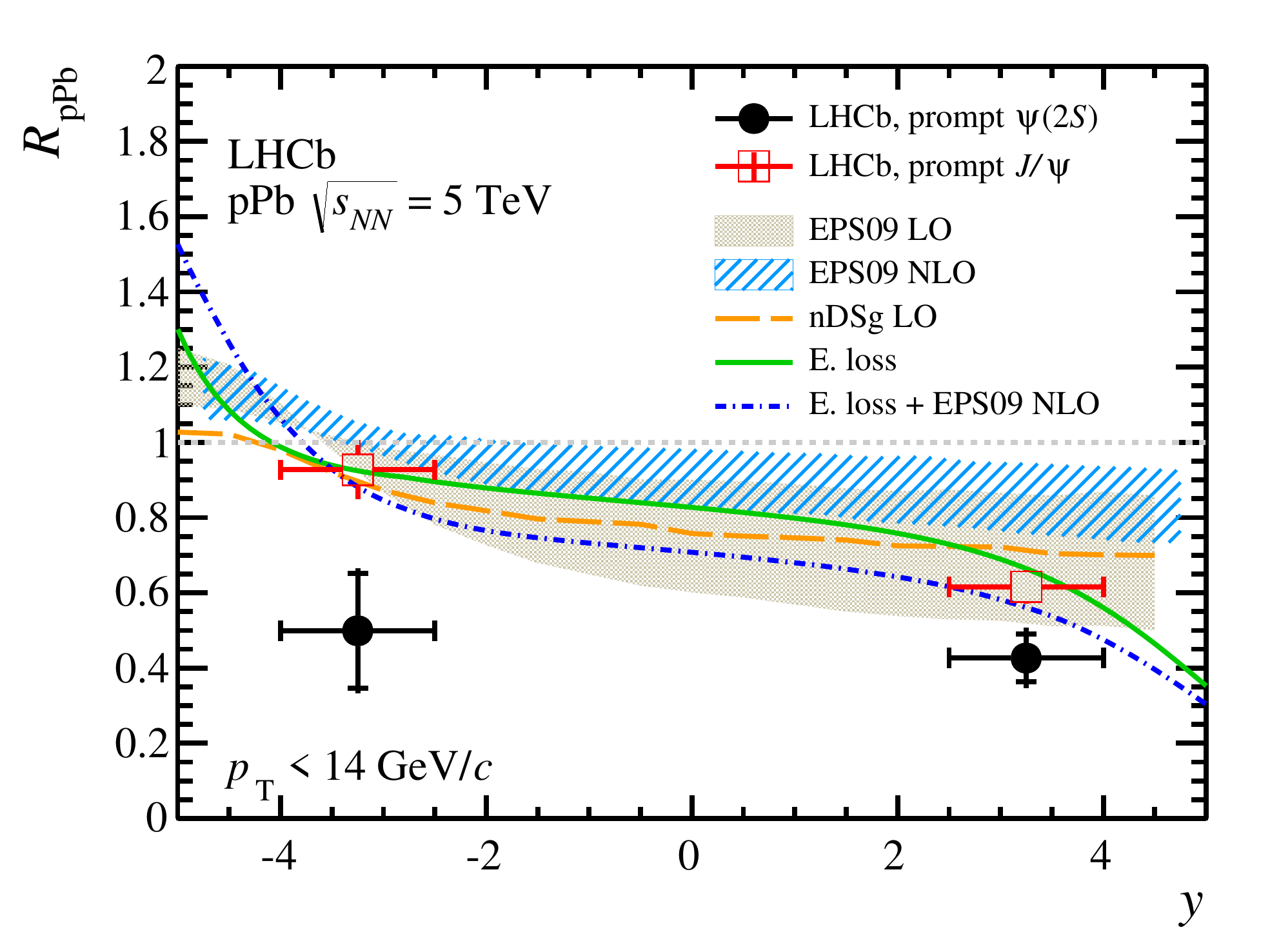}}
\end{minipage}
\hfill
\begin{minipage}{0.32\linewidth}
\centerline{\includegraphics[width=1.0\linewidth]{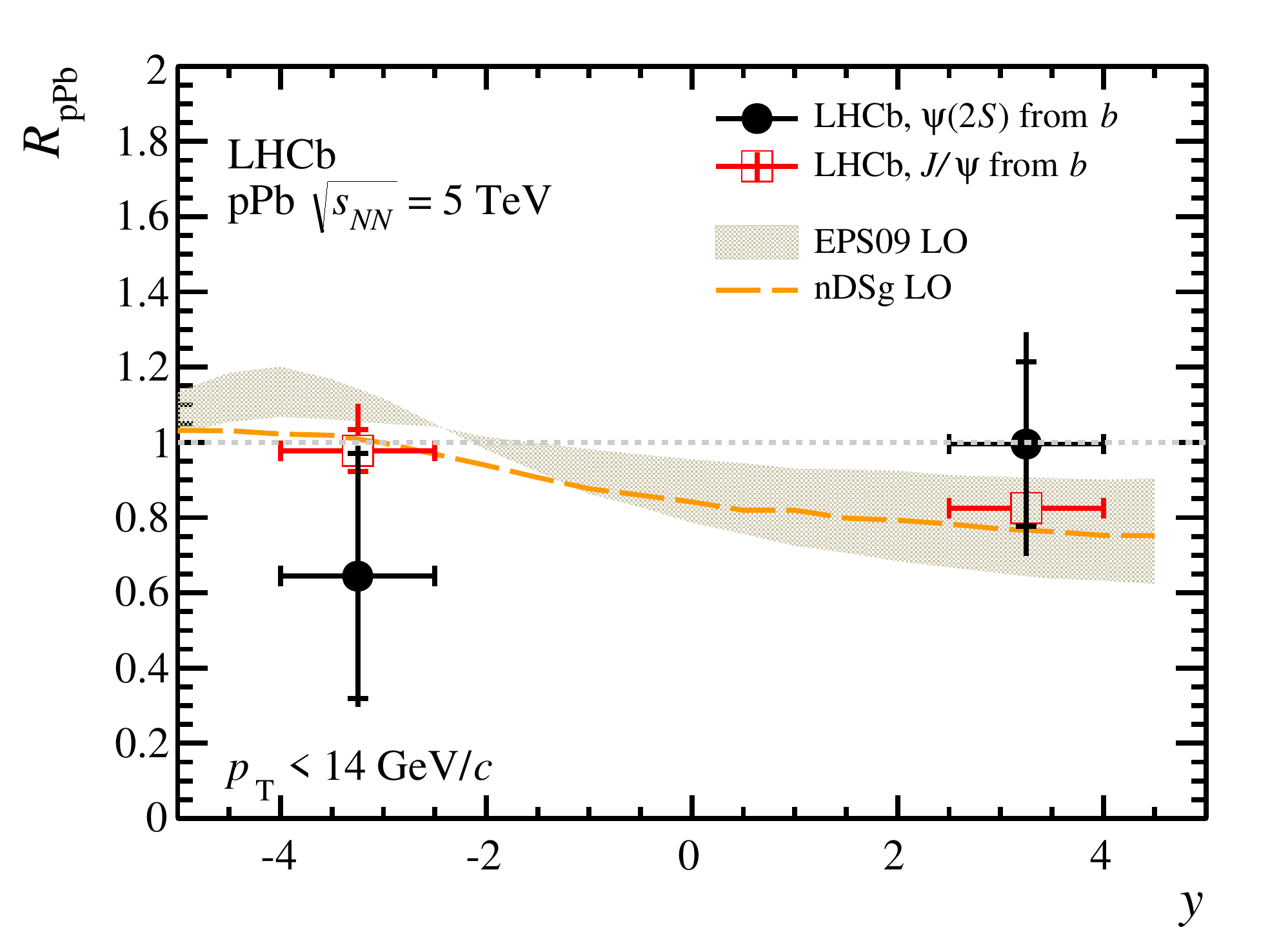}}
\end{minipage}
\hfill
\begin{minipage}{0.32\linewidth}
\centerline{\includegraphics[width=1.0\linewidth]{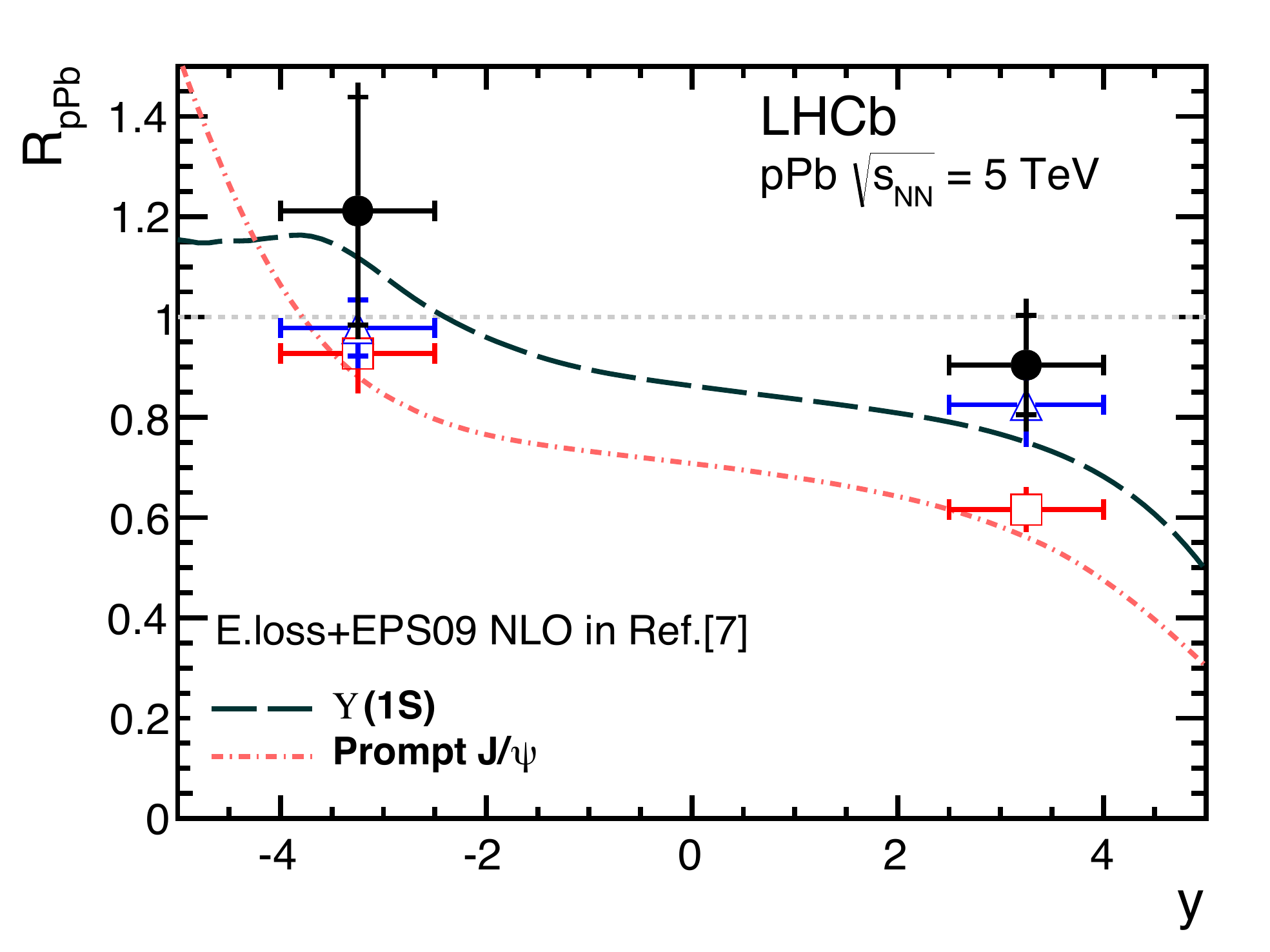}}
\end{minipage}
\caption[]{The nuclear modification factor $R_{p\mathrm{Pb}}$ in bins of $y$ 
    integrated over $p_\mathrm{T}$ for (left) prompt $J/\psi$ and $\psi(2S)$, (middle) $J/\psi$-, $\psi(2S)$-from-$b$ and (right) $\Upsilon(1S)$. }
\label{fig:RpPbQuarkonia}
\end{figure}

LHCb also studied the CNM effects in $p$Pb data using the productions of quarkonia, including $J/\psi$, $\psi(2S)$ and
$\Upsilon$ mesons~\cite{JpsiInPA,PsiInPA,UpsilonInPA}. The prompt $J/\psi$ and $\psi(2S)$
mesons are separated from those from $b$-hadron decays, allowing
to study the CNM effects for both components. The results of $R_{p\mathrm{Pb}}$ for the quarkonium states are given in
Fig.~\ref{fig:RpPbQuarkonia}. It can be seen that in the forward sample, $J/\psi$ production in $p$Pb is strongly
suppressed compared to that in $pp$, the suppression for $\psi(2S)$ is even stronger, while that for $\Upsilon(1S)$ is
modest. In the backward data sample, $R_{p\mathrm{Pb}}$ is compatible with unity for $J/\psi$ and $\Upsilon$ mesons, but
intriguing strong suppression is seen for $\psi(2S)$, suggesting that the mechanism for $\psi(2S)$ production in $p$Pb
collisions is not the same as that for $J/\psi$, which is to be understood. 
In the forward sample,  $R_{p\mathrm{Pb}}$ for $\psi$-from-$b$ is closer to unity than for prompt $\psi$ mesons, which
indicates that the open bottom hadrons in $p$Pb are less suppressed compared to prompt charmonium. The measurements of $R_{p\mathrm{Pb}}$ for $J/\psi$
and $\Upsilon$ are in good agreements with various theoretical calculations referred to in the corresponding paper.

\section{Prospects of LHCb heavy ion studies}
At LHCb, the fixed target program benefits from the SMOG device. 
When noble gas is injected in SMOG, the proton and lead beams can collide with the gas similar to fixed target collisions. 
With different choices of noble gases, we are able to explore different sizes of colliding systems. 
In the past, LHCb has already collected short runs of SMOG collisions, including
$p$Ne at $\sqrt{s_\mathrm{NN}}=110$ GeV, and PbNe at $\sqrt{s_\mathrm{NN}}=54$ GeV, and clear $J/\psi$ signals are
observed in these samples. LHCb also collected PbPb collisions in 2015, and more will be available in the coming years. 

\section{Summary}
In conclusion, heavy flavors are good tools to study the physics in heavy ion collisions, and LHCb has demonstrated its
capabilities to contribute significantly to heavy ion studies. 
LHCb has studied cold nuclear matter effects using quarkonia and open charm productions with the $p$Pb data at 
$\sqrt{s_\mathrm{NN}}=5$ TeV.
LHCb also collected PbPb collisions, allowing to study rich physics programs covering heavy flavours, electroweak, soft
QCD and QGP physics. LHCb is also unique to do fixed target physics,
exploiting  colliding systems of different sizes at low energies using the SMOG system.

\section*{Acknowledgments}
The corresponding author acknowledges support from the 
European Research Council (ERC) through the project EXPLORINGMATTER,
funded by the ERC through a ERC-Consolidator-Grant.

\end{document}